\documentclass[12pt,preprint2]{aastex}

\def\pc{\,{\rm pc}}
\def\kpc{\,{\rm kpc}}
\def\etal{{et al.}}
\def\eg{{\it e.g.}}

\def\ie{{\it i.e.}}

\def\kms{$\mathrm {km}~\mathrm{s}^{-1}$}
\def\degrees{^\circ}
\def\spose#1{\hbox to 0pt{#1\hss}}
\def\gtsim{\mathrel{\spose{\lower.5ex \hbox{$\mathchar"218$}}
     \raise.4ex\hbox{$\mathchar"13E$}}}
\def\ltsim{\mathrel{\spose{\lower.5ex\hbox{$\mathchar"218$}}
     \raise.4ex\hbox{$\mathchar"13C$}}}

\slugcomment{Accepted to ApJ}
\lefthead{Debattista \etal}
\righthead{The Kinematic Signature of Face-On Peanut-Shaped Bulges}

\begin{document}

\title{The Kinematic Signature of Face-On Peanut-Shaped Bulges}

\author{Victor P.~Debattista\altaffilmark{1,}\altaffilmark{2},
C. Marcella Carollo} \affil{Institut f\"ur Astronomie, ETH Z\"urich,
CH-8093, Z\"urich, Switzerland} \email{debattis@astro.washington.edu,
marcella.carollo@phys.ethz.ch}

\author{Lucio Mayer \& Ben Moore}
\affil{Department of Theoretical Physics, University of Z\"urich,
  Winterthurerstrasse 190, CH-8057, Z\"urich, Switzerland}
\email{lucio@physik.unizh.ch, moore@physik.unizh.ch}

\altaffiltext{1}{current address: Astronomy Department, University of
Washington, Box 351580, Seattle WA, 98195}
\altaffiltext{2}{Brooks Fellow}

\begin{abstract}
  We present a kinematic diagnostic for peanut-shaped bulges in nearly
  face-on galaxies.  The face-on view provides a novel perspective on
  peanuts which would allow study of their relation to bars and disks
  in greater detail than hitherto possible.  The diagnostic is based
  on the fact that peanut shapes are associated with a flat density
  distribution in the vertical direction.  We show that the kinematic
  signature corresponding to such a distribution is a minimum in the
  fourth-order Gauss-Hermite moment $s_4$.  We demonstrate our method
  on $N$-body simulations of varying peanut strength, showing that
  strong peanuts can be recognized to inclinations $i \simeq
  30\degrees$, regardless of the strength of the bar.  We also
  consider compound systems in which a bulge is present in the initial
  conditions as may happen if bulges form at high redshift through
  mergers.  We show that in this case, because the vertical structure
  of the bulge is not derived from that of the disk, that the
  signature of a peanut in $s_4$ is weakened.  Thus the same kinematic
  signature of peanuts can be used to explore bulge formation
  mechanisms.  The observational requirements for identifying peanuts
  with this method are challenging, but feasible.
\end{abstract}


\section{Introduction}

About $25\%$ of the stellar luminosity in the universe comes from the
bulges of disk galaxies (Persic \& Salucci 1992; Fukugita et
al. 1998).  Therefore understanding how bulges form is a necessary
step in understanding galaxy formation in general.

If bulges are distinct entities, rather than just disk light in excess
of an exponential (van den Bosch \etal\ 2002; B\"oker \etal\ 2003), a
mechanism for generating them separate from disk formation must be
considered.  Because bulges sit at the bottom of the potential wells
of galaxies, many paths for their formation are possible.  Bulge
formation scenarios can be classified loosely based on whether the
driving mechanism is internal or external.  A widely discussed example
of externally-driven bulge formation is in the merger at early times
of dwarf-sized galactic subunits around which disks subsequently grow
(Kauffmann \etal\ 1993).  Observational evidence supporting this
scenario includes the relatively homogeneous bulge stellar populations
in the Milky Way (Ferreras \etal\ 2003; Zoccali \etal\ 2003) and
Andromeda (Stephens \etal\ 2003), and counter-rotation found in some
galaxies (Pizzella \etal\ 2004).

Discussion of internally driven bulge formation has focused on the
secular evolution of disk instabilities.  Observational evidence
supporting secular bulge formation includes disk-like,
almost-exponential light profiles (Andredakis \& Sanders 1994;
Courteau \etal\ 1996; de Jong 1996; Carollo \etal\ 2001; Carollo 1999;
Carollo \etal\ 1998; MacArthur \etal\ 2003), occasionally disk-like,
cold kinematics (Kormendy 1993; Kormendy \etal\ 2002), the correlation
between the scale-lengths of bulges and disks (de Jong 1996; MacArthur
\etal\ 2003) and the similar average colors of bulges and inner disks
(Terndrup \etal\ 1994; Peletier \& Balcells 1996; Courteau \etal\
1996).  The recent review of Kormendy \& Kennicutt (2004) summarizes
our current understanding of, and evidence for, secular formation of
some bulges.

The bulges of many edge-on galaxies are box- or peanut- (B/P) shaped
(Burbidge \& Burbidge 1959; Jarvis 1986).  Binney \& Petrou (1985)
constructed axisymmetric models of B/P bulge systems including
cylindrical rotation as observed (Kormendy \& Illingworth 1982).  They
speculated that accretion is responsible for creating such systems;
however observations found little evidence of accretion onto them
(Shaw 1987; Whitmore \& Bell 1988, but see also L\"utticke \etal\
2004).  A different scenario emerged from 3-D $N$-body simulations,
namely formation via secular evolution of bars (Combes \& Sanders
1981), either through resonant scattering or through bending
instabilities (Pfenniger 1984; Combes \etal\ 1990; Pfenniger \&
Friedli 1991; Raha \etal\ 1991).  The orbits supporting peanuts have
been studied extensively (Combes \etal\ 1990; Pfenniger 1984;
Pfenniger 1985; Patsis \etal\ 2002a) and shown to generally arise from
vertically unstable x1 orbits.  Patsis \etal\ (2002b) showed that
these orbits are present and peanuts are possible even if the
non-axisymmetric perturbation is very weak.

Thereafter, observational efforts sought to establish the connection
between B/P-shaped bulges and bars by seeking evidence for a bar in
edge-on B/P-bulged systems.  In the case of NGC 4442, the B/P bulge is
already apparent at an inclination of $72\degrees$, at which the bar
also can be recognized (Bettoni \& Galletta 1994).  A second such case
is NGC 7582 at an inclination of $65\degrees$ (Quillen \etal\ 1997).
In several B/P bulges, photometric features of a bar have been claimed
(\eg\ de Carvalho \& da Costa 1987) but the bar interpretation is not
unique when only photometry is available.  The fraction of edge-on
bulges having B/P shapes is $\sim 45\%$ (L\"utticke \etal\ 2000),
which is consistent with the fraction of galaxies containing bars
($\sim 70\%$, Knapen \etal\ 2000; Eskridge \etal\ 2000) once the
arbitrary orientations of bars to the line-of-sight (LOS) are
considered.  However, the most important evidence for the presence of
bars in B/P bulges comes from a comparison of the edge-on gas and
stellar LOS velocity distributions (LOSVDs) of $N$-body bars (Bureau
\& Athanassoula 1999; Athanassoula \& Bureau 1999; Bureau \&
Athanassoula 2004) and real galaxies (Kuijken \& Merrifield 1995;
Merrifield \& Kuijken 1999; Bureau \& Freeman 1999; Chung \& Bureau
2004).

These edge-on studies have established the connection between
B/P-shaped bulges and bars.  However, the degeneracy inherent in
deprojecting edge-on galaxies makes it difficult to study other
properties of the host galaxy.  Moreover, while B/P shapes are
produced by bars, this does not exclude the possibility that bulges
are {\it shaped} by secular processes, not formed by them.  Addressing
this issue requires an attempt at a cleaner separation of bulges, bars
and peanuts.  In face-on systems the viewing geometry is
well-constrained and bars are readily apparent.  If we can also
recognize peanuts in them then we obtain an important new perspective
on the relation of peanuts and bars.  For example, this permits study
of the relative sizes of bars and peanuts: meager observational
evidence suggests that these need not be equal (Kormendy and Kennicutt
2004), in agreement with simulations (below).  Moreover, for
inclinations $\sim 30\degrees$, it becomes possible to measure
accurately the pattern speed of the bar (Debattista 2003) and
therefore to test for resonances and compare with theoretical
predictions.  It would also allow determination of the fraction of
barred galaxies with peanuts, which may be different from the fraction
of peanuts with bars.  And finally, as we will show below, the ability
to detect peanuts face-on opens the possibility of exploring bulge
formation mechanisms.

In this paper we examine the kinematic signature of peanuts in face-on
galaxies.  In Section \ref{sec:exact} we first explore some simple
analytic models to help understand the behavior of more realistic
systems.  Working with Gauss-Hermite moments (Gerhard 1993; van der
Marel \& Franx 1993), we show that the fourth-order LOSVD moment,
$s_4$, is monotonically increasing with $d_4$, the fourth-order
vertical density moment.  Thus $s_4$ can be used to probe the vertical
structure of a disk.  We describe the $N$-body building in Section
\ref{sec:ics} and in Section \ref{sec:nbody} we present the $N$-body
models, with and without strong peanuts, used in this paper and
examine their vertical density distributions.  We show that the main
signature of a peanut is in $d_4$, rather than in the disk
scale-height.  We explore the vertical LOSVDs of these $N$-body models
in Section \ref{sec:kine}, showing that $s_4$ can be used as a robust
kinematic signature of a peanut, independent of bar strength.  In
Section \ref{sec:inc} we show that moderate inclinations do not
substantially degrade the diagnostic.  Section \ref{sec:observs}
discusses the required signal-to-noise and spectral resolution and our
conclusions are presented in Section \ref{sec:discussion}.

\section{Exact results}
\label{sec:exact}

We first consider some exact models useful for interpreting the
results of $N$-body simulations.  Peanuts constitute a density
distribution more vertically extended than the surrounding disk.  Let
us denote the root-mean-square (RMS) height and vertical velocity as
$h_z$ and $\sigma_z$ respectively.  Deviations from Gaussian
distributions can be parameterized by the moments of an expansion in
Gauss-Hermite functions (Gerhard 1993; van der Marel \& Franx 1993).
The second order term in such an expansion is related to the RMS.  The
third order term measures deviations which are asymmetric about the
mean and are therefore likely to be small for the vertical density and
velocity distributions of disk galaxies.  The fourth order term
measures the lowest order symmetric deviation from a Gaussian; it is
negative when a distribution is broader than Gaussian and positive
when it is more peaked.  We denote the fourth order Gauss-Hermite
moment of the vertical density distribution as $d_4$ and that for the
LOSVD as $s_4$.  Following Gerhard (1993), for a vertical LOSVD
$l(v_z)$ normalized to the projected surface density $\Sigma$, we
define
\begin{equation}
s_4 = \frac{\sqrt{4 \pi}}{\Sigma}\int l(w) H_4(w) e^{-\frac{1}{2}w^2} dw
\end{equation}
where $w = (v_z-\overline{v_z})/\sigma_z$ and $H_4(w) =
\frac{1}{\sqrt{768 \pi}}(16 w^6 - 48 w^2 + 12)$.  A similar expression
holds for $d_4$.  For a particle model, the integral becomes a sum and
$\Sigma$ is replaced by $N_p$, the number of particles in a bin.

The vertical density extension associated with the peanut will
correspond to an increase in $h_z$ (the ``scale-height'') and/or a
decrease in $d_4$.  What is the observable effect on the LOSVD of such
variations?  First consider how $\sigma_z$ varies as a function of
radius in the case where $h_z$ is constant.  In a single-component
axisymmetric system, the one-dimensional vertical Boltzmann$+$Poisson
equation is
\begin{equation}
\frac{\partial}{\partial z} \left( \frac{1}{\rho}
\frac{\partial}{\partial z} \rho \overline{v_z^2} \right) = - 4 \pi G \rho
\label{eqn:sigmaz}
\end{equation}
(\eg\ Binney \& Tremaine Eqn. 4-38).  If the system is isothermal,
then $\overline{v_z^2}$ is independent of $z$ (and is therefore equal
to $\sigma_z^2$).  The solution of Eqn. \ref{eqn:sigmaz} is
\begin{equation}
\rho(z) = \rho_0 \ {\rm sech}^2 (z/z_0)
\end{equation}
(Spitzer 1942), where $\rho_0$ is the density in the mid-plane and
\begin{equation}
z_0 = \frac{\sigma_z}{\sqrt{2 \pi G \rho_0}}
    = \frac{\sigma_z^2}{\pi G \Sigma}.
\label{eqn:z0}
\end{equation}
(Note that for the isothermal disk, $h_z = z_0 \pi/\sqrt{12}$.)  Thus
if $z_0$ is (nearly) constant (as suggested by observations [van der
Kruit \& Searle 1981; de Grijs \& Peletier 1997]) then $\sigma_z^2
\propto \Sigma$.  Since typically $\Sigma(R) \propto e^{-R/R_{\rm
d}}$, then $\sigma_z \propto e^{-R/2R_{\rm d}}$ and it is likely that
any signature of the peanut in $\sigma_z$ will be swamped by this
density-driven radial variation.

\begin{figure}[!ht]
  \plotone{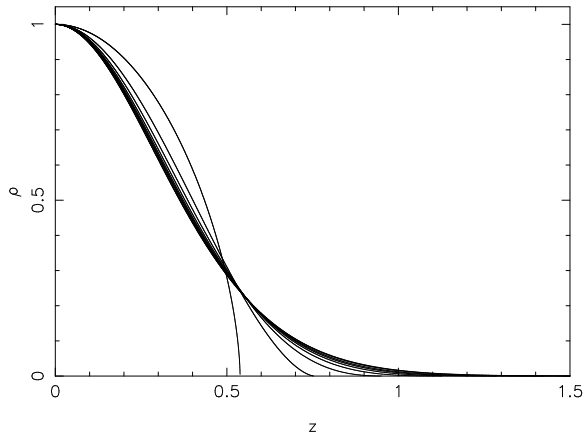}
\caption{ Model III of Camm (1950).  Various vertical density
  profiles, with $n$ increasing in order of increasing maximum $z$,
  from $n=1.6$ in steps of 1, are shown. 
\label{fig:cammdens}}
\end{figure}

\begin{figure}[!ht]
  \plotone{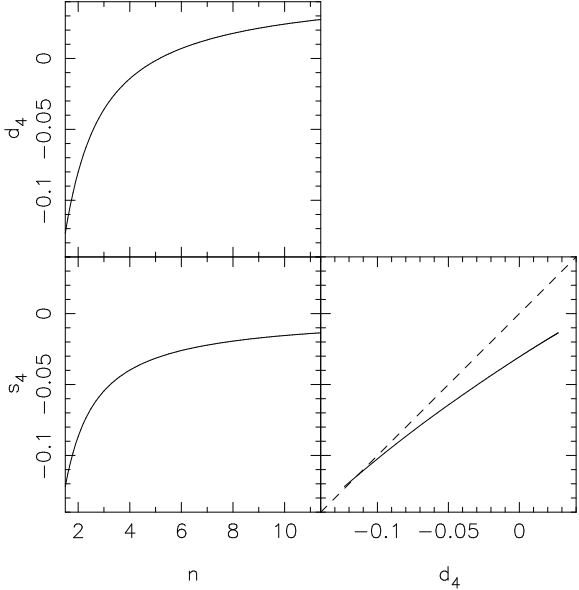}
\caption{ The Gauss-Hermite moments $d_4$ and $s_4$ of Camm's (1950)
  model III.  The left panels show the variation of these parameters
  with $n$, while the panel on the right plots $s_4$ versus $d_4$.
  The dashed diagonal line indicates $s_4 = d_4$
\label{fig:cammz4h4}}
\end{figure}

Next we use simple models to explore the signature in the LOSVD of a
flat-topped vertical density distribution.  Consider Camm's (1950)
series of analytic solutions of the collisionless Boltzmann equation
for systems stratified in plane-parallel layers of infinite extent.
This is a good local approximation to real galaxies at low $z$ when
the rotation curve is flat (van der Kruit \& Freeman 1986).
In Camm's model III, the density distribution is given by $\rho(z) =
\rho_0 \cos^{2-2/n} \theta$, where $n > \frac{3}{2}$, and the
parameter $\theta$ is defined by the relation
\begin{equation}
z = A \int_0^\theta\sec^{1-2/n}\phi~ d\phi
\end{equation}
with $A$ some constant.  Several examples of this density distribution
for different $n$ are presented in Figure \ref{fig:cammdens}.  The
corresponding distribution function is
\begin{equation}
f(w,\theta) = C(n) (2 n \cos^{2/n}\theta - w^2)^{n - \frac{3}{2}},
\end{equation}
where $C(n)$ is a normalization constant.  These densities and
distribution functions can be integrated numerically to compute $d_4$
and $s_4$.  The results are presented in Figure \ref{fig:cammz4h4}.
The limit $n \rightarrow \infty$ corresponds to the isothermal sheet,
in which the distribution function has the same Gaussian dependence on
velocity at all heights.  Thus in this limit, $s_4 = 0$.  On the other
hand, the isothermal sheet has a ${\rm sech}^2 z$ profile which is
more peaked than a Gaussian and therefore has $d_4 > 0$.  At smaller
$n$, the density profile becomes increasingly flat-topped leading to
$d_4 < 0$ which drives $s_4 < 0$.  Two properties of $s_4$ make it an
excellent probe of $d_4$.  First is the fact that $s_4$ increases
monotonically with $d_4$, which makes $s_4$ an observable surrogate
for the unobservable $d_4$.  Second, $s_4 \ltsim d_4$, so that the
vertical velocity distribution is generally broader than the density
distribution, which makes it observationally robust.

\section{$N$-body Systems}
\label{sec:ics}

We use $N$-body models with different initial conditions and evolved
on different codes to study the kinematic signatures of face-on
peanuts.  Since gas is generally depleted within bars, and moreover
dissipates its vertical energy, it is not a good tracer of face-on
peanuts.  We therefore focus on only the stellar kinematics of
peanuts, and our simulations are all collisionless.  Table
\ref{tab:sims} lists all the simulations used in this study.

\subsection{Rigid-halo simulations}

The highest resolution simulations in this paper were run on a 3-D
cylindrical particle-mesh (PM) grid code (described in Sellwood \&
Valluri 1997).  The main advantages of this code for the present study
are that it permits high spatial and mass resolutions; indeed these
simulations used $\geq 4M$ disk particles and force softening
$\epsilon \leq 2 z_{\rm d}/3$, where $z_{\rm d}$ is the (constant)
Gaussian vertical scale-height of the initial conditions.  Since
Gauss-Hermite moments generally require high signal-to-noise ratios to
be measured reliably, a large number of particles is desirable.  At
the same time, the high force resolution ensures that the vertical
motions of particles are well resolved.

\begin{table*}[!ht]
\begin{centering}
\begin{tabular}{ccccccccccc}\hline 
\multicolumn{1}{c}{Run} &
\multicolumn{1}{c}{$N_*$} &
\multicolumn{1}{c}{$n$} &
\multicolumn{1}{c}{$R_{\rm d}/\epsilon$} &
\multicolumn{1}{c}{$z_{\rm d}/\epsilon$} &
\multicolumn{1}{c}{$Q$} &
\multicolumn{1}{c}{$f_{\rm b}$} &
\multicolumn{1}{c}{Halo} &
\multicolumn{1}{c}{$r_{\rm h}$} &
\multicolumn{1}{c}{$v_{\rm h}$} &
\multicolumn{1}{c}{Peanut} \\ \hline 
R1 &        $7.5M$ & 1.0 & 60 & 3.0 & 1.2 & 0.0 & Log. & 3.3 & 0.68 &  Strong   \\ 
R2 &     $7.5M$ & 1.0 & 60 & 6.0 & 1.2 & 0.0 & Log. & 3.3 & 0.68 &  Strong   \\
R3 &        $4.0M$ & 1.0 & 60 & 3.0 & 2.4 & 0.0 & Hern. & 20.8 & 1.44 &  Strong \\ 
R4 &        $7.5M$ & 2.5 & 60 & 3.0 & 1.0 & 0.0 & Log. & 3.3 & 0.68 &  Strong   \\ 
R5 &        $7.5M$ & 1.0 & 60 & 6.0 & 2.4 & 0.0 & Log. & 3.3 &  0.68 &  Weak    \\ 
R6 &        $7.5M$ & 1.0 & 60 & 12. & 1.2 & 0.0 & Log. & 3.3 & 0.68 &  None     \\ 
R7 &        $4.0M$ & 1.0 & 60 & 3.0 & 1.6 & 0.0 & Hern. & 20.8 & 1.44 &  None   \\ 
R8 &        $7.5M$ & 1.0 & 60 & 3.0 & 1.2 & 0.0 & Log. & 3.3 & 0.68 &  None     \\ \hline

B1 &        $4.0M$ & 1.0 & 80 & 8.0 & 2.5 & 0.2 & Log. & 5.0 & 0.65 &  Weak     \\ 
B2 &        $4.0M$ & 1.0 & 80 & 8.0 & 1.9 & 0.2 & Log. & 5.0 & 0.65 &  Strong     \\
B3 &        $4.0M$ & 1.0 & 80 & 8.0 & 1.3 & 0.2 & Log. & 5.0 & 0.65 &  Strong   \\ \hline

 L1    &            $0.2M$ & 1.0 & 40 & 6.0 & 0.0 & 0.0 & NFW & 108 & 76 & Strong   \\ \hline 
\end{tabular}
\caption{The sample of simulations used in this paper.  $N_*$ is the
number of disk$+$bulge particles, $n$ is the index of the initial
S\'ersic disk, $R_{\rm d}$, $z_{\rm d}$ and $\epsilon$ are the
scale-length, scale-height of the {\it initial} disk and softening
length, $Q$ is the initial disk Toomre-$Q$ and $f_{\rm b}$ is the
bulge mass as a fraction of the total (disk$+$bulge).  In column
``Halo'' we describe the type of halo used: logarithmic, Hernquist or
NFW.  $r_{\rm h}$ and $v_{\rm h}$ are the halo scale-length and
characteristic velocity, respectively.  In column ``Peanut'' we give a
qualitative description of the peanut: strong, weak or none.  For the
live-halo system L1, we give the minimum $Q$, $z_{\rm d}$ is for a
${\rm sech}^2$ profile, $r_{\rm h} = r_{vir}$ and $v_{\rm h} =
V_{vir}$ in \kpc\ and \kms\ respectively.}
\label{tab:sims}
\end{centering}
\end{table*}

The rigid halos were represented by either a spherical logarithmic
potential,
\begin{equation}
\Phi_L(r) = \frac{v_{\rm h}^2}{2}~ \ln(r^2 + r_{\rm h}^2),
\end{equation}
or a Hernquist (1990) model
\begin{equation}
\Phi_H(r) = -\frac{M_{\rm h}}{r+r_{\rm h}}.
\end{equation}
where $r_{\rm h}$ is a halo scale-radius, $v_{\rm h}$ is a
characteristic halo velocity and $M_{\rm h}$ is a halo mass.  We
define $v_{\rm h} \equiv \sqrt{GM_{\rm h}/r_{\rm h}}$ for the
Hernquist halos.

The initially axisymmetric disks were all S\'ersic type
\begin{equation}
\rho_{\rm d}(R,z) \propto (1-f_{\rm b})M e^{-(R/R_{\rm d})^{1/n}}
e^{-\frac{1}{2}(z/z_{\rm d})^2}
\end{equation}
where $f_{\rm b}$ is the fraction of the active (\ie\ bulge$+$disk)
mass which is in the bulge, $M$ is the active mass, $R_{\rm d}$ is the
disk scale-length, $z_{\rm d}$ is the Gaussian thickness and $n$ is
the S\'ersic index ($n=1$ corresponding to an exponential profile and
$n=4$ to a de Vaucouleurs profile).  Disk kinematic setup used the
epicyclic approximation with constant Toomre-$Q$ and the vertical
Jeans equation to set vertical motions appropriate for a constant
Gaussian thickness.  We use units where $R_{\rm d} = M = G = 1$, which
gives a unit of time $(R_{\rm d}^3/GM)^{1/2}$.

Bulges were generated using the method of Prendergast \& Tomer (1970),
where a distribution function is integrated iteratively in the global
potential until convergence.  We used the isotropic distribution
function of a lowered polytrope, truncated at $r_{\rm b}$
\begin{equation}
f(\vec{x},\vec{v}) \propto \left[-E(\vec{x},\vec{v}) \right]^{1/2} - 
          \left[-E_{\rm max}\right]^{1/2}.
\label{eqn:df}
\end{equation}
Here $E_{max} = \Phi_{tot}(r_{\rm b})$, the total potential at $r_{\rm
  b}$ in the disc plane.  For all bulges we set $r_{\rm b} = 0.78$.
  The bulges in runs B2 and B3 were fully rotating, while that in run
  B1 had no rotation.  Further details of the compound system setup
  method used can be found in Debattista \& Sellwood (2000).

The polar grids were $N_R\times N_\phi \times N_z = 60 \times 64
\times 225$ or larger.  For all the PM simulations, the vertical
spacing of the grid planes, $\delta z$, was set to $0.0125$.  We used
Fourier terms up to $m=8$ in the potential, which was softened with
the standard Plummer kernel.  Time integration was performed with a
leapfrog integrator with a fixed time-step, $\delta t = 0.02$ for
simulations B1-B3, $\delta t = 0.0025$ for run R4 and $\delta t =
0.01$ for all the rest.

\subsection{Live-halo simulation}
\label{ssec:livehalos}

The disadvantage of the PM code is that we needed to use a rigid halo.
Therefore in run L1 we used a lower mass resolution live-halo
simulation run with {\sc pkdgrav} (Stadel 2001), a multi-stepping,
parallel treecode.

The live-halo model was built using the technique developed by
Hernquist (1993; see also Springel \& White 1999).  We start with an
isotropic NFW halo (Navarro, Frenk \& White 1996) with virial radius,
$R_{vir} = 108$ \kpc, circular velocity at the virial radius, $V_{vir}
= 76$ \kms\ and virial mass $M_{vir} = 1.5 \times 10^{11} M_\odot$.
Then adiabatic contraction of the halo due to the presence of the disk
is taken into account assuming that the spherical symmetry of the halo
is retained and that the angular momentum of individual dark matter
orbits is conserved (see Springel \& White 1999).  The disk mass
fraction relative to the halo virial mass, $f_d=M_{\rm d}/M_{vir} =
0.08$.  We used an exponential disk with scale-length $R_{\rm d} =
1.99$ \kpc\ and a ${\rm sech}^2(z/z_d)$ vertical profile.  We set
$z_d = 0.15 R_{\rm d}$ and softening length $\epsilon = 50$ \pc.  The
velocity field of the disk was calculated as in Springel \& White
(1999) assuming the radial and vertical velocity dispersions are
equal, $\sigma_R = \sigma_z$, with $\sigma_R$ chosen to give minimum
Toomre-$Q = 1.2$.  Then the azimuthal velocity dispersion is
determined from $\sigma_R$ using the epicyclic approximation.

\section{Vertical Density of the $N$-body Models}
\label{sec:nbody}

The evolution of model B1 has been described in Debattista (2003),
while runs B2 and B3 formed part of the preliminary survey for the
Milky Way modeling described in Bissantz \etal\ (2004).  The evolution
of most of the remaining models will be described elsewhere
(Debattista \etal\ 2005 in progress).  Here we are interested
primarily in the final systems not in details of their evolution.
Except for run R3, which formed only a very weak oval distortion, all
these simulations formed bars.  The edge-on view of the simulations
R1-R8 is presented in Figure \ref{fig:peanuts}.  Throughout this
paper, we use a convention where the bar is along the $x$-axis and the
$z$-axis is perpendicular to the disk.  Runs R1-R4 contain prominent
peanuts which were produced by bending instabilities (Raha \etal\
1991; Merritt \& Sellwood 1994).  To better present these peanuts, in
Figure \ref{fig:pntslice} we present the edge-on projected density of
particles in the narrow range $-0.5 \leq y \leq 0.5$, \ie\ we show
only a narrow slice of each model extending to about the minor-axis of
the bar.  This gives a better appreciation of the peanuts which will
be sought in the face-on view, where the disks do not mask peanuts.
Peanuts can form in weakly barred systems (Patsis \etal\ 2002); the
peanut in run R3 formed in the presence of only a very weak oval.  As
a result, this peanut is almost axisymmetric.  Run R5 contains a weak
peanut while runs R6-R8 have no peanuts at all.  Run R8 is identical
to run R1 except for one important detail: we forced symmetry about
the mid-plane.  Therefore, although the velocity ellipsoid is very
anisotropic with $\sigma_z/\sigma_R \simeq 0.25$ through most of the
bar, the system could not bend and did not develop a peanut.

Comparing the peanuts visible in Figures \ref{fig:peanuts} and
\ref{fig:pntslice} with the bars seen in the face-on surface density
overlaid on Figure \ref{fig:heights}, it is apparent that the peanut
in run R1 is smaller then the bar.  Kormendy and Kennicutt (2004)
noted that the peanut in the moderately inclined galaxy NGC 7582 was
significantly shorter than the bar, and worried that this may be a
problem for a secular peanut formation scenario.  Run R1 shows that
peanuts need not fill the entire major axis of a bar.  In contrast,
the peanut in run R2 extends to about the ends of the bar.

Figure \ref{fig:heights} shows maps of $h_z$ for runs R1-R8.  On the
bar's major axis $h_z$ increases radially outwards, reaching a maximum
when a peanut is present.  In most cases, the local maximum in $h_z$
occurs close to the peanut.  On the other hand, in runs R6-R8, none of
which contain a peanut, $h_z$ increases throughout.

The maps in Figure \ref{fig:z4} show $d_4$ for runs R1-R8.  Minima in
$d_4$ correlate well with the location of peanuts and there are no
significant minima in $d_4$ other than at $R=0$ in the absence of a
peanut.

\begin{figure}[!ht]
  \plotone{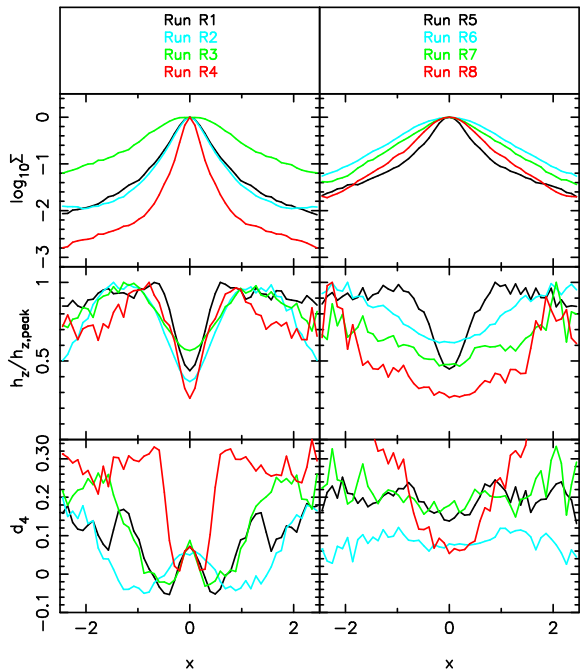}
\caption{A summary of the face-on density moments along the bars'
major axis for runs R1-R8.
\label{fig:photmjraxis}}
\end{figure}

For better comparison of the different models, Figure
\ref{fig:photmjraxis} shows the density, $h_z$ and $d_4$ profiles
along the major axes of the face-on bars.

\begin{figure*}[!ht]
\caption{The edge-on view of runs R1-R8.  In all cases, the bar has
been rotated into the $x$-axis, so it is viewed side-on.
\label{fig:peanuts}}
\end{figure*}

\begin{figure*}[!ht]
\caption{The edge-on view of the systems as in Figure
\ref{fig:peanuts} but with $|y| \leq 0.5$ only shown.
\label{fig:pntslice}}
\end{figure*}

\begin{figure*}[!ht]
\caption{The face-on view of runs R1-R8 showing the color-coded RMS
height ($h_z$) of particles.  The contours show the projected surface
density, $\Sigma$.
\label{fig:heights}}
\end{figure*}

\begin{figure*}[!ht]
\caption{The face-on view of runs R1-R8 showing the fourth-order
Gauss-Hermite moment of the vertical density distribution, $d_4$.
\label{fig:z4}}
\end{figure*}

\subsection{$h_z$ versus $d_4$}

Are the second and fourth density moments equivalent ways of defining
the peanut?  Figure \ref{fig:photmjraxis} shows that the maxima in
$h_z$ corresponds very well to the minima in $d_4$ in runs R1 and R2.
In run R3 the peak of $h_z$ is at a different radius from the minimum
of $d_4$, while in run R4 $h_z$ and $d_4$ have {\it maxima} at about
the same point.  Thus the maximum in $h_z$ and the minimum in $d_4$
are not equivalent ways of defining the peanut.

Which of $h_z$ and $d_4$ is the better tracer of peanuts?  Figure
\ref{fig:photmjraxis} shows that the major-axis $h_z$ profiles of runs
R1 and R5 are rather similar, suggesting very similar peanuts; Figures
\ref{fig:peanuts} and \ref{fig:pntslice} show that this is far from
being the case.  In contrast, their $d_4$ profiles in Figure
\ref{fig:photmjraxis} are very different.  For this reason $d_4$ is a
better measure of the presence and strength of a peanut.  The reason
why $h_z$ is not an optimal peanut diagnostic is that it is partly
determined by the local projected density, as suggested by the
correlation, evident in Figure \ref{fig:photmjraxis}, between the
depth of the central minimum in $h_z$ and the central concentration.

\section{Vertical Kinematics of Peanuts}
\label{sec:kine}

\subsection{The absence of a peanut signature in $\sigma_z$}
\label{ssec:disp}

Figure \ref{fig:sigz} shows maps of $\sigma_z$ for runs R1-R8.  No
sign of the peanuts is evident in these maps.  We found that, in the
region $-2.5 \leq x,y \leq 2.5$ of all the models, $\sigma_z$
correlates very strongly with $\Sigma$, even off the bar's major axis.
The signature of a peanut is buried in the small scatter in $\sigma_z$
at fixed $\Sigma$, making the peanut hard to distinguish from
$\sigma_z$.

\begin{figure*}[!ht]
\caption{The face-on view of runs R1-R8 showing $\sigma_z$.
\label{fig:sigz}}
\end{figure*}

\subsection{A peanut signature in $\sigma_z^2/\Sigma$}

\begin{figure}[!ht]
  \plotone{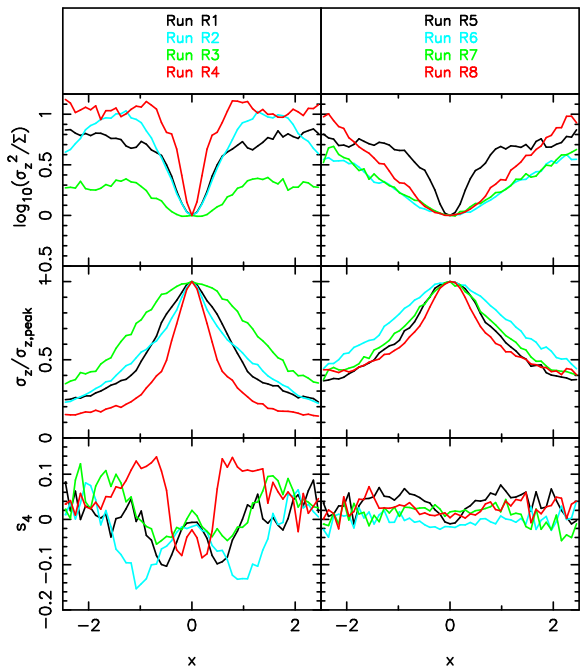}
\caption{A summary of the face-on kinematic moments along the bar's
major axis for runs R1-R8.  The peanut is not apparent in the velocity
dispersion profile but is prominent in the $s_4$ profile.
\label{fig:kinemjraxis}}
\end{figure}

Following the discussion in Section \ref{sec:exact}, it is
unsurprising that we cannot identify peanuts from $\sigma_z$ profiles.
Since $\sigma_z^2 \sim \Sigma$ in the isothermal disk
(Eqn. \ref{eqn:z0}), it is worth exploring whether $\sigma_z^2/\Sigma$
is better than $\sigma_z$ at locating peanuts.  In Figure
\ref{fig:kinemjraxis} we plot, for runs R1-R8, $\sigma_z^2/\Sigma$ on
the bars' major-axes.  In all cases, the profiles of
$\log(\sigma_z^2/\Sigma)$ are rather similar to those of $h_z$
(although there is not a one-to-one correspondence).  However, a
comparison of Figures \ref{fig:photmjraxis} and \ref{fig:kinemjraxis}
shows that the minimum in $d_4$, which we showed above to be a good
indicator of a peanut, does not correspond to any special point in the
$\sigma_z^2/\Sigma$ profile.  Therefore $\sigma_z^2/\Sigma$ has the
same limitations as a peanut diagnostic as $h_z$, although it may
still be useful in distinguishing between peanut and peanutless
systems.

\subsection{The peanut signature in $s_4$}
\label{ssec:h4}

Figure \ref{fig:h4} plots $s_4$ for runs R1-R8.  A qualitative
difference between the peanut systems and the peanutless ones is
evident --- two negative minima at the location of the peanut on the
bar's major axis if a peanut is present.  In runs R1-R4 there is a
considerable variation in bar strength, but in each case, the negative
$s_4$ minimum criterion recognizes the peanut, demonstrating that it
does not depend on bar strength.

In run R5, which produced a weak peanut, $s_4$ remains greater than
zero on the bar major-axis and the only minimum is at $R=0$.  Thus the
kinematic diagnostic cannot identify weak peanuts.  Other than at
$R=0$, no significantly negative minimum in $s_4$ occurs in runs R6-R8
which lack a peanut.

\begin{figure*}[!ht]
\caption{The face-on view of runs R1-R8 showing $s_4$.
\label{fig:h4}}
\end{figure*}

\begin{figure*}[!ht]
  \plotone{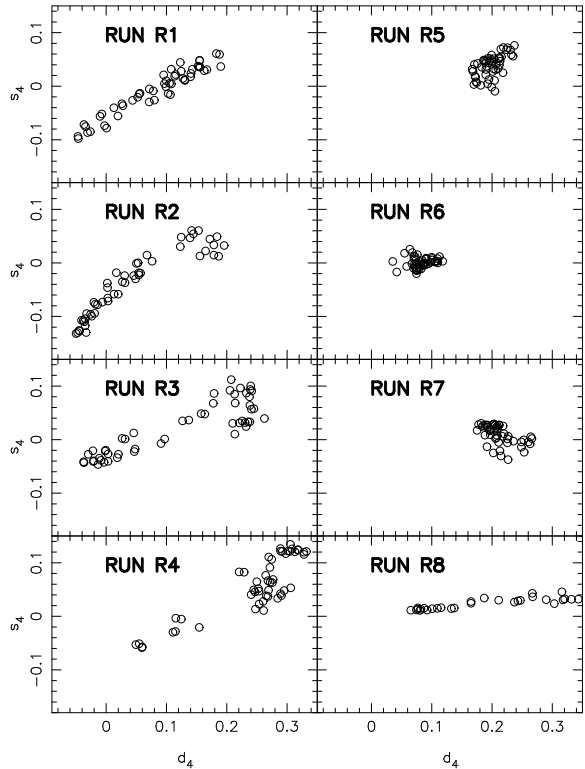}
\caption{The variation of $s_4$ versus $d_4$ on the bar's major axes
for runs R1-R8.
\label{fig:h4vsz4}}
\end{figure*}

Figure \ref{fig:h4vsz4} plots $s_4$ versus $d_4$ on the major axes of
the bars.  As predicted by the analytic models of Section
\ref{sec:exact}, the minimum in $d_4$, which is a tracer of the
peanut, corresponds to the minimum of $s_4$.  Therefore minima in
$s_4$ are an excellent kinematic peanut diagnostic in these face-on
systems.

\subsection{Live-halo simulation}
\label{ssec:live}

Mainly because of the lower mass resolution, which results in
significantly lower $S/N$ in the Gauss-Hermite moments, we used the
live-halo run L1 only to confirm results of the rigid-halo
simulations.  Despite the lower $S/N$, we were still able to identify
clear and well-matched minima in $d_4$ and $s_4$ in the region of the
peanut.  Indeed the properties of the peanut in L1 are very similar to
those in run R1 although the initial conditions, including vertical
structure, were quite different.  This gives us confidence that the
peanut diagnostic developed from rigid-halo simulations is not an
artifact of the rigid halos.

\subsection{Simulations with bulges}
\label{ssec:bulges}

Now we consider simulations which include a bulge in the initial
conditions.  In Figure \ref{fig:bulges} we plot the edge-on views of
these systems.  A weak peanut is present in run B1, which is masked in
the full edge-on view.  A stronger peanut is present in run B2 and an
even stronger one in B3.  Comparing with Figures \ref{fig:peanuts} and
\ref{fig:pntslice}, the pure disk components of runs B1, B2 and B3 are
most like runs R5, R4 and R1 respectively, albeit only approximately.
Figure \ref{fig:bulgemaps} plots maps of $d_4$ and $s_4$ and the major
axis profiles are presented in Figure \ref{fig:bulgpprof}.  In all
three runs, the slope of the profile of $h_z$ has a break which allows
the peanut to be recognized.  Thus the peanuts are still evident in
$\sigma_z^2/\Sigma$, including again the weak peanut in B1.  All three
models have minima in their $d_4$ profiles, but those in runs B1 and
B2 are broad ones extending down to $R=0$.  In $s_4$, no minima are
visible in run B1 (and thus no peanut is identified, as was the case
also for the weak peanut in run R5), while the usual minima
identifying a peanut are clear in run B3.  The case of run B2 is more
interesting.  If the bulge in this system were dark, a peanut would
stand out clearly in $s_4$ (cyan line); with the addition of the
bulge, the different bulge kinematics, especially the significantly
higher $\sigma_z$, perturb the net $s_4$, hiding the presence of a
peanut in the sense that only a single broad minimum down to $R=0$
remains in its profile.

\begin{figure}[!ht]
\caption{The disk$+$bulge simulations.  Panels (a) show the bulges,
panels (b) the disks and panels (c) the bulges$+$disks. Panels (d) and
(e) show the bulges+disks and disks in the range $|y| \leq 0.5$.
\label{fig:bulges}}
\end{figure}

\begin{figure*}[!ht]
\caption{The bulge simulations B1-B3.
\label{fig:bulgemaps}}
\end{figure*}

\begin{figure}[!ht]
\caption{The photometric profiles of runs B1-B3 on the bars' major
axis.  Red (dotted) is the bulge, cyan (dashed) is the disk and black
(solid) is disk+bulge.
\label{fig:bulgpprof}}
\end{figure}

\begin{figure}[!ht]
  \plotone{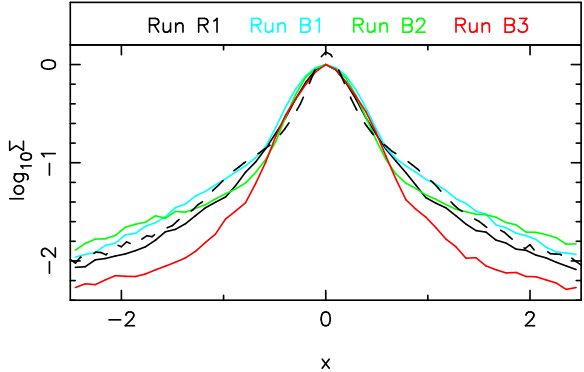}
\caption{A comparison of the density profiles along the bar's major
axis in runs R1 and B1-B3.  The central concentration is largely due
to initial conditions in runs B1-B3 but results from secular evolution
in run R1.  The dashed black line shows the scaled-profile along the
bar major-axis of NGC 4477.
\label{fig:compbulgedisk}}
\end{figure}

\section{The Effect of Inclination}
\label{sec:inc}

Exactly face-on galaxies are rare; the probability for a galaxy to be
within $5\degrees$ of edge-on is over 20 times larger than for it to
be within $5\degrees$ of face-on ($8.7\%$ versus $0.4\%$).  The
inclination needs to be within $24\degrees$ of face-on before its
probability is equal to that for $5\degrees$ from edge-on.  A sample
of exactly face-on galaxies may therefore be hard to obtain.  Thus it
is necessary to ask what happens to the kinematic signature of a
peanut when it is viewed not quite perfectly face-on.  Is it possible
that the negative minimum in $s_4$ signature of a peanut is erased, or
induced where no peanut is present, for other than an exactly face-on
orientation?

Once a system is no longer perfectly face-on, the observed LOS
velocity dispersion, $\sigma_{\rm los}$, relative to which $s_4$ is
defined, includes contributions from the radial ($\sigma_R$) and
tangential ($\sigma_\phi$) dispersion components.
If $i$ is the inclination angle, $\phi$ is any angle in the disk's
plane measured relative to the inclination axis, $\alpha \equiv
\sigma_\phi/\sigma_R $ and $\beta \equiv \sigma_z/\sigma_R$, then the
contribution of $\sigma_z$ to $\sigma_{\rm los}$ is
\begin{equation}
\left(\frac{\sigma_z}{\sigma_{\rm los}}\right)^2 = \frac{1}{\beta^{-2}
\sin^2 \phi \sin^2 i + \alpha^2 \beta^{-2} \cos^2 \phi \sin^2 i +
\cos^2 i}.
\end{equation}
This has a maximum along the disk's major axis ($\phi = 0$) and a
minimum on the minor axis ($\phi = 90 \degrees$).  A crude estimate of
$\sigma_z/\sigma_{\rm los}$ can be obtained assuming that $\alpha^2 =
1/2$ (\ie\ a flat rotation curve).  Then if $\beta = 0.293$ (the
minimum value required for stability [Araki 1985]), we estimate that
the contamination may be as high as $10\%$ already at an inclination
of $10\degrees$.  However, $\beta = 0.293$ is extreme: in the Solar
neighborhood, Dehnen \& Binney (1998) find $\beta = 0.53 \pm 0.07$
while Gerssen \etal\ (2000) find even larger $\beta$ in earlier Hubble
types.  The contamination of $\sigma_z$ in $\sigma_{\rm los}$ is still
less than $10\%$ at $i = 30\degrees$ on the disk's major axis if
$\beta = 0.53$.

\begin{figure}[!ht]
  \plotone{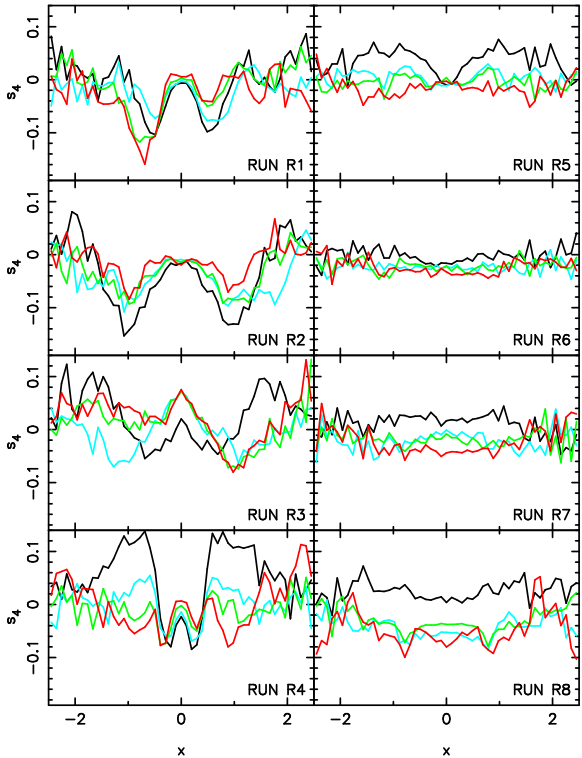}
\caption{The $s_4$ profiles on the bar major-axis seen face-on (black
  line) and at $i=30\degrees$ for runs R1-R8.  The bar makes an angle
  $\phi = 0\degrees$, $45\degrees$ and $90\degrees$ for the cyan,
  green and red lines, respectively.  In our notation $x>0$ is the
  side nearer the observer.
\label{fig:inclined}}
\end{figure}

Since these estimates are based on simplifying approximations, we also
explored the effect of inclination directly on the $N$-body
simulations to $i=40\degrees$.
Figure \ref{fig:inclined}, presents $s_4$ on the bar major-axis for
runs R1-R8 inclined at $i = 30\degrees$ and with the bar oriented at
$0\degrees \leq \phi \leq 90\degrees$.

For strong peanuts, the negative $s_4$ minimum criterion still
distinguishes between peanut and peanutless systems up to an
inclination of $\sim 30\degrees$.  When a peanut is present, the two
minima on opposite sides of the bar become asymmetric as $\phi$
increases to $90\degrees$.  Therefore $\phi \ltsim 45\degrees$, is a
more favorable orientation for finding peanuts.  In some instances,
inclination produces negative minima in $s_4$ off the bar's major axis
in both peanut and peanutless systems.  However, inclination leads to,
at most, only shallow minima in peanutless systems, although an
overall negative $s_4$ can result when no peanut is present.  (Thus a
negative $s_4$ without a minimum is not by itself sufficient as a
peanut diagnostic.)

In run R8 we prevented bending by forcing mid-plane symmetry, which
resulted in a final $\sigma_z/\sigma_R \simeq 0.25$.  When this system
is viewed at an inclination of $30\degrees$, two {\it shallow} minima
in $s_4$ appear at all $\phi$ (see Figure \ref{fig:inclined}) even
though no peanut is present.  As expected from the analytic estimate
above, inclination has a much larger effect on the vertical kinematic
moments of vertically cold systems.  As this is an unrealistically
anisotropic system (symmetrization about the disk plane having
inhibited the bending instability), this simulation represents an
extreme extent to which inclination introduces minima in $s_4$ when no
peanut is present.

Two animations accompany the online version of this paper.  These show
the effect of inclination on the $s_4$ moments of simulations B1 and
B3.

\section{Observational Requirements}
\label{sec:observs}

Measuring kinematic Gauss-Hermite moments requires high $S/N$ spectra
(Bender \etal\ 1994).  In our $N$-body measurements, this has been
possible because of the large number of particles.  By resampling
experiments, we determined that a $S/N \gtsim 50$ is required to
measure $s_4$ sufficiently accurately to identify a peanut, in good
agreement with Bender \etal\ (1994).  Fortunately bars are generally
bright features which helps improve the $S/N$.

The kinematic signature of a peanut is strong for some distance on the
bar's minor-axis (Figure \ref{fig:h4}).  Thus very precise placement
of the slit along the bar's major axis is not necessary.  This also
allows the widest slit consistent with the necessary spectral
resolution.

The spectral resolution, {\sc R}, required depends on the value of
$\sigma_z$ which varies from galaxy to galaxy.  The Milky Way has
$\sigma_z \sim 100$ \kms\ (Kuijken 2003).  Thus one would need {\sc R}
$\simeq 2500$ to find a peanut in a face-on galaxy like the Milky Way.
If, on the other hand, $\sigma_z \sim 30$ \kms\ (\eg\ Bottema 1993),
{\sc R} $\simeq 8500$ would be needed.

\section{Discussion}
\label{sec:discussion}

B/P-shaped bulges are common in edge-on disks and simulations show
that these can form by secular evolution in barred galaxies.  In this
paper we explored the signature of peanuts on the vertical density and
the resulting kinematics.  We showed that both $h_z$ and $d_4$ are
affected by a peanut, but the two are not equivalent signatures of a
peanut.  The preferred peanut signifier is $d_4$ since it
distinguishes between weak and strong peanuts.

The vertical velocity dispersion, $\sigma_z$ is a poor diagnostic for
peanuts because it depends on the local density.  The quantity
$\sigma_z^2/\Sigma$, which in an isothermal disk would trace $h_z$,
factors out some of this dependence and is able to identify peanuts,
even weak ones, at breaks in its slope.  However it is unable to
quantify peanut strength and correlates poorly with $d_4$.  This
parameter may also be prone to systematic effects from variations in
mass-to-light ratios if, for example, the bulge and the disk are
composed of different stellar populations, as would happen if the
bulge formed at high redshift through mergers.  Nevertheless, this is
a useful peanut diagnostic that is worth testing in real galaxies.

An excellent kinematic diagnostic of face-on peanuts are negative
double minima in the Gauss-Hermite moment $s_4$.  The negative $s_4$
minimum signature of a peanut holds for any bar strength down to the
weakest of ovals and may therefore be used to search also for peanuts
in unbarred galaxies.  This diagnostic is not too sensitive to
inclination for $i<30\degrees$, with negative minima on a bar's
major-axis continuing to be associated with the presence of a peanut.
However, inclination leads to an asymmetry between the two sides of
the bar as the bar orientation approaches the minor axis.  Thus a bar
oriented within $\sim 45\degrees$ of the line-of-nodes is ideal for a
peanut search.

\subsection{Bulge formation mechanisms and vertical structure}

Two competing models of bulge formation --- via internal secular
evolution and via external drivers --- each account for a significant
body of observational evidence, suggesting that both processes play
some role (\eg\ Wyse 2004).  Thus we need to ask which process
dominates in which galaxies.  The results here suggest a novel
observational program to address this question by targeting the degree
of decoupling of the vertical structure of bulges and bars.  To be
concrete, consider runs R4 and B2, both of which contain a peanut.
Since the bulge is already present in the initial conditions in run B2
(as would be a bulge formed in an early merger), the presence of a
peanut cannot be used to address whether the bulge formed by secular
evolution or not.  Photometrically the two have similar density
profiles (Figure \ref{fig:compbulgedisk}) since run R1 acquired a
central density concentration by secular evolution (Hohl 1971;
Debattista \etal\ 2004).  These profiles are typical of the major-axis
profiles of real galaxies: the dashed line in this figure shows the
$J$-band profile of the nearly face-on galaxy NGC 4477, taken from the
online\footnote{Available at http://www.astro.princeton.edu/$\sim$frei/catalog.htm } Frei catalog of galaxies (Frei \etal\ 1996).
However the vertical structure of the bulge in run B2 is qualitatively
different from that in run R1 because it is not derived from the
disk's.  Because of this, neither the $d_4$ nor the $s_4$ profiles
have the kind of separated minima associated with peanuts in which the
central concentration forms purely by secular evolution.  Peanuts have
been shown to be visible in $45\%$ of edge-on galaxies which means
that they are even more common since projection hides some fraction of
peanuts.  Thus a kinematic survey of face-on barred galaxies should
turn up a large fraction of galaxies with the negative $s_4$ minima
signature of a peanut if bulges are built largely by secular evolution
of bars.  On the other hand, if bulges formed largely through mergers
of dwarfs then peanuts would need to be stronger to be identifiable in
face-on kinematics.

\subsection{The effect of gas}

The simulations presented here were all collisionless.  If gas
funneled by a bar plays an important role also in bulge formation,
naively it would seem likely that the vertical kinematic signature of
a peanut becomes confused.  However, in a barred potential, gas sinks
to small radii where its kinematics and of stars formed from it do not
perturb those from the peanut further out.  Lower resolution
hydro+$N$-body live-halo experiments we have run and which will be
presented elsewhere show that peanuts can still be recognized in that
case with the same stellar kinematic diagnostic.  Therefore an
observational survey of face-on barred galaxies to look for peanuts
appears worth undertaking.

\acknowledgments 

We would like to thank the anonymous referee for comments which helped
improve the presentation of this paper.  V.P.D. thanks Enrico Maria
Corsini, Sven De Rijcke and Ortwin Gerhard for fruitful discussion.

\end{document}